\documentclass{aa}  
\usepackage{graphicx}
\usepackage{txfonts}
\usepackage[colorlinks=true]{hyperref}
\hypersetup{linkcolor=blue, citecolor=blue}
\usepackage[draft]{todonotes}
\usepackage{amsmath,systeme}
\usepackage{multicol}

\usepackage{soul}
\usepackage{xcolor}
\usepackage{amsmath}

\begin{document} 

   \title{High-accuracy estimation of magnetic field strength in the interstellar medium from dust polarization}

    \titlerunning{High-assuracy B-field estimation from dust polarization}
   
   \author{Raphael Skalidis  \and Konstantinos Tassis}
    
   \institute{Institute of Astrophysics, Foundation for Research and Technology-Hellas, Vasilika Vouton, GR-70013 Heraklion, Greece\\
            \email{rskalidis@physics.uoc.gr}, \and 
            Department of Physics \& ITCP, University of Crete, GR-70013, Heraklion, Greece \\
            \email{tassis@physics.uoc.gr} 
             }

    \authorrunning{Skalidis \& Tassis}
   \date{}

 
  \abstract
   {A large-scale magnetic field permeates our Galaxy and is involved in a variety of astrophysical processes such as star formation and cosmic ray propagation. Dust polarization has been proven one of the most powerful observables for studying the field properties in the interstellar medium (ISM). However, it does not provide a direct measurement of its strength. Different methods have been developed which employ both polarization and spectroscopic data in order to infer the field strength. The most widely applied method has been developed by \cite{davis_1951} and \cite{chandra_fermi} (DCF). 
   The DCF method relies on the assumption that isotropic turbulent motions initiate the propagation of Alfvén waves. Observations, however, indicate that turbulence in the ISM is anisotropic and non-Alfvénic (compressible) modes may be important.}
   {Our goal is to develop a new method for estimating the field strength in the ISM, which includes the compressible modes and does not contradict the anisotropic properties of turbulence.}
   {We adopt the following assumptions: 1) gas is perfectly attached to the field lines; 2) field line perturbations propagate in the form of small-amplitude MHD waves; 3) turbulent kinetic energy is equal to the fluctuating magnetic energy. We use simple energetics arguments that take into account the compressible modes to estimate the strength of the magnetic field.}
   {We derive the following equation: $B_{0}=\sqrt{2 \pi\rho} \delta v /\sqrt{\delta \theta}$, where $\rho$ is the gas density, $\delta v$ is the rms velocity as derived from the spread of emission lines, and $\delta \theta$ is the dispersion of polarization angles. We produce synthetic observations from 3D MHD simulations and we assess the accuracy of our method by comparing the true field strength with the estimates derived from our equation. We find a mean relative deviation of $17 \%$. The accuracy of our method does not depend on the turbulence properties of the simulated model. In contrast DCF, even when it is combined with the \cite{hildebrand_2009} and \cite{houde_2009} method, systematically overestimates the field strength.}
   {Compressible modes can affect significantly the accuracy of methods that are based solely on Alfvénic modes. The formula that we propose includes compressible modes; however it is applicable only in regions with no self-gravity. Density inhomogeneities may bias our estimates to lower values.}
   
             
   \maketitle
%

\newcommand{\Sturb}{$\sigma_{v}$}
\newcommand{\dangles}{$\delta \theta$}
\newcommand{\Su}{$\sigma_{u}$}
\newcommand{\Bordered}{$\vec{B_{0}}$}
\newcommand{\Bturb}{$\vec{B_{t}}$}
\newcommand{\MA}{$M_{A}$}
\newcommand{\MS}{$M_{s}$}
\newcommand{\VA}{$\rm{V_{A}}$}
\newcommand{\HI}{H$_{\rm{I}}$}
\newcommand{\CO}{$^{12}$CO}

\section{Introduction}

The Galactic magnetic field plays a key role in various astrophysical processes, such as cosmic ray propagation and star formation, and affects foregrounds relevant to cosmic microwave background polarization experiments. Various tracers exist revealing either the line of sight (LOS) or the plane of the sky (POS) component of the field. Zeeman splitting is sensitive to the LOS component and is the only observable, which reveals both the LOS local magnitude and the orientation of the field vector along this direction. Regarding the POS component, synchroton and dust polarization are the most common observables. Unlike synchroton, dust polarization traces the POS field in the cold neutral medium, where stars form. Dust polarization reveals only the morphology of the field and not its strength \citep[see][for a recent review]{andersson_review}.  Different magnetic field strength estimation methods have been developed using the dust polarization data.

The first method was presented by \cite{davis_1951} and \cite{chandra_fermi} (DCF). They assumed that magnetic field lines are distorted due to the propagation of the incompressible transverse magnetohydrodynamic (MHD) waves, known as Alfvén waves. This distortion induces spread in the polarization angle distribution, which, combined with the gas turbulent motions from spectroscopic data, allows the estimate of the true magnetic field strength. Different effects have been recognized to bias the accuracy of the DCF method towards higher values \citep{zweibel_1990, myers_1991}. \cite{ostriker_2001}, \cite{padoan_2001}, and \cite{heitsch_2001}, after testing the accuracy of the method in MHD simulations, found that on average the method produces a two-fold deviation. In addition, it was realized that external forces, like self-gravity, can bend the field lines and induce extra dispersion in the polarization angle distribution. In order to treat the problem, \cite{girart_2006} fitted parabolas to the polarization data and removed the large-scale hour-glass bending from the polarization data of a pre-stellar core. A similar, but more sophisticated approach, was followed by \cite{pattle_2017}. On the other hand, \cite{hildebrand_2009} and \cite{houde_2009} (HH09) developed an analytical model for the polarization data, which measures the turbulence-induced spread in the presence of any external source of B-field bending. Modifications of the DCF method have been developed by \cite{heitsch_2001}, \cite{kudoh_basu_2003}, \cite{falceta_2008}, \cite{cho_2016}, \cite{yoon_cho_2019}, and \cite{lazarian_2020}.  See \cite{pattle_2019} and \cite{hull_zhang_2019} for recent reviews. \textit{All these methods rely on the assumption that the Alfvén waves are producing the observed polarization angle dispersion and the linewdiths in the emission spectra.}

The interstellar medium (ISM) is highly compressible and other than the Alfvén waves, contains MHD wave modes that induce density compressions. These are known as fast and slow magnetosonic modes and their existence in astrophysical plasmas is inevitable because they are excited by the Alfvén waves, e.g. \cite{Heyvaerts_1983}. In addition, the so-called entropy modes can contribute to the observed compressibility of the ISM \citep{lithwick_2001}, however they produce zero velocity and magnetic field fluctuations. All the aforementioned methods ignore the existence of the compressible modes. This can lead to significant inaccuracies in the magnetic field strength estimates. 

In the present work, we propose a new relation for estimating the field strength, which, unlike the DCF method, \textit{includes the compressible modes}. We assess the validity of this method in synthetic observations that we produce from 3D MHD turbulence simulations. Our method employs both polarization and spectroscopic data. Its applicability should be restricted to regions that do not show large-scale bending due to self-gravity.

The structure of the paper is as follows. In Section~\ref{sec:dcf_method} we critically review the DCF method. We review the underline assumptions and pont out projection effects that may affect its accuracy. We test the method in 3D simulations. In Section~\ref{sec:hdcf_method} we critically review the HH09 method and we apply it in synthetic observations of 3D simulations. In Section~\ref{sec:proposed_method} we present our new method and apply it in 3D simulations and discuss its limitations. In Section~\ref{sec:conclusions} we summarize our results.


\section{Classical DCF method}
\label{sec:dcf_method}

\subsection{Foundations of the method}

We decompose the total magnetic field into a mean, $\vec{B_{0}}$, and a fluctuating component $\vec{\delta B}$. The total field is $\vec{B} = \vec{B_{0}} + \vec{\delta B}$ with a total magnetic energy density equal to
\begin{equation}
    \label{eq:magnetic_energy}
    \frac{B^{2}}{8\pi} = \frac{1}{8\pi} [B_{0}^{2} + \delta B^{2} + 2 \vec{\delta B} \cdot \vec{B_{0}}] , 
\end{equation}
where bold letters are used to denote vectors. The last two terms correspond to changes of the magnetic energy, $\delta \epsilon_{m}$, due to $\vec{\delta B}$ fluctuations. DCF assumed that the ISM plasma conductivity is infinite. This means that the magnetic field is "frozen-in" the gas, hence both gas and field lines oscillate in phase. Turbulent gas motions perturb the field lines and initiate small amplitude fluctuations, $|\vec{\delta B}| \ll |\vec{B_{0}}|$, \textit{in the form of Alfvén waves} about the mean field. DCF assumed that the kinetic energy of turbulent motions will be equal to the fluctuating magnetic energy density
\begin{equation}
    \frac{1}{2}\rho \delta v^{2} = \frac{\delta B^{2}}{8\pi} ,
\end{equation}
where $\rho$ is the gas density and $\delta v$ the rms velocity. Note that $\vec{B_{0}} \cdot \vec{\delta B} = 0$, since Alfvén waves are transverse. We divide both sides by $B_{0}^{2}$ and after rearranging we obtain
\begin{equation}
    B_{0} = \sqrt{4 \pi \rho} \delta v \Bigg [ \frac{\delta B}{{B_{0}}} \Bigg ]^{-1}.
\end{equation}

The magnetic field orientation is traced by dust polarization (with a $\pi$ ambiguity) and the dispersion of the polarization angle distribution, \dangles, is a metric of $\delta B/B_{0}$. If the mean field is stronger than the fluctuating component, the field lines will appear approximately straight, hence \dangles\ will be small. If, on the other hand, the fluctuations are relatively large, field lines will be dispersed by turbulent motions and \dangles\ will increase. Thus, DCF assumed $\delta \theta = \delta B/B_{0}$, yielding 
\begin{equation}
    \label{eq:dcf_derivation_dtheta_approximation}
    B_{0} = \sqrt{\frac{4 \pi \rho}{3}} \frac{\delta v}{ \delta \theta }, 
\end{equation}
where the factor $1/ \sqrt{3}$ was inserted by DCF because they assumed that turbulent motions are isotropic and only one of the three Cartesian velocity components perturbs the field lines. 
Other authors \citep[see][]{ostriker_2001} proposed a different correction factor $f$. The generalized DCF equation is then 
\begin{equation}
    \label{eq:classical_dcf}
    B_{0} = f \sqrt{4 \pi \rho}\frac{\delta v}{\delta \theta} .
\end{equation}
The mean magnetic field, $B_{0}$, can be written in velocity units by dividing by $\sqrt{4 \pi \rho}$, thus
\begin{equation}
    \label{eq:classical_dcf_speed}
    V_{A} = f \frac{\delta v}{\delta \theta}, 
\end{equation}
where \VA\ is the Alfvén speed. 

\subsection{Caveats of the method}

The DCF approach is based on the assumption of equipartition between kinetic and magnetic energy, which holds for travelling MHD waves. For standing waves the total energy oscillates between magnetic and kinetic forms. Since our observables (polarization angles and spectroscopic data) are instantaneous, equipartition between kinetic and magnetic energies can only happen in standing waves twice in a phase cycle. Standing waves have been produced in MHD simulations, e.g. \citep{kudoh_basu_2003} and have been identified in the ISM, e.g. in the Musca molecular cloud \citep{tritsis_2018}. In this paper we consider only travelling MHD waves.

\subsubsection{Turbulent velocities and compressible modes}
\label{sec:turbulent_velocities}

The DCF method has been used extensively in atomic and molecular clouds. Turbulent velocities, denoted as $\delta v$ or $\sigma_{v}$, are measured using spectroscopic data, e.g. \HI\ $21$cm line, CO(J=1-0) line, etc. Emission lines are approximated as Gaussians and non-thermal linewidths are usually observed. The non-thermal broadening, $\sigma_{v, turb}$, is attributed to  turbulent gas motions, so
\begin{equation}
    \sigma_{v, turb}^{2} = \sigma_{v, tot}^{2} - \sigma_{v, thermal}^{2},
\end{equation}
where $\sigma_{v, tot}$ is the total observed spread and $\sigma_{v, thermal}$ the thermal broadening.


Turbulent broadening, $\sigma_{v, turb}$, may contain contributions from wave modes other than the Alfvén modes. MHD plasma also supports the propagation of fast and slow modes, which can be excited even if they are not initially in the system \citep{Heyvaerts_1983}, due to their coupling with Alfvén modes. These modes can induce extra dispersion in the observed velocities and significantly affect the DCF method, which neglects their contribution. 

The ISM is highly compressible \citep[e.g.,][]{heiles_2003}. This implies that $\sigma_{v, turb}$ includes velocities from both Alfvén and compressible modes. As a result, $\sigma_{v, turb}$ will always be higher than by Alfvén waves alone, hence $B_{0}$ will be overestimated. This makes the mode decomposition necessary in order to apply the DCF method accurately. However, mode decomposition is not trivial in observations.

    \begin{figure*}
    	\includegraphics[width=\hsize]{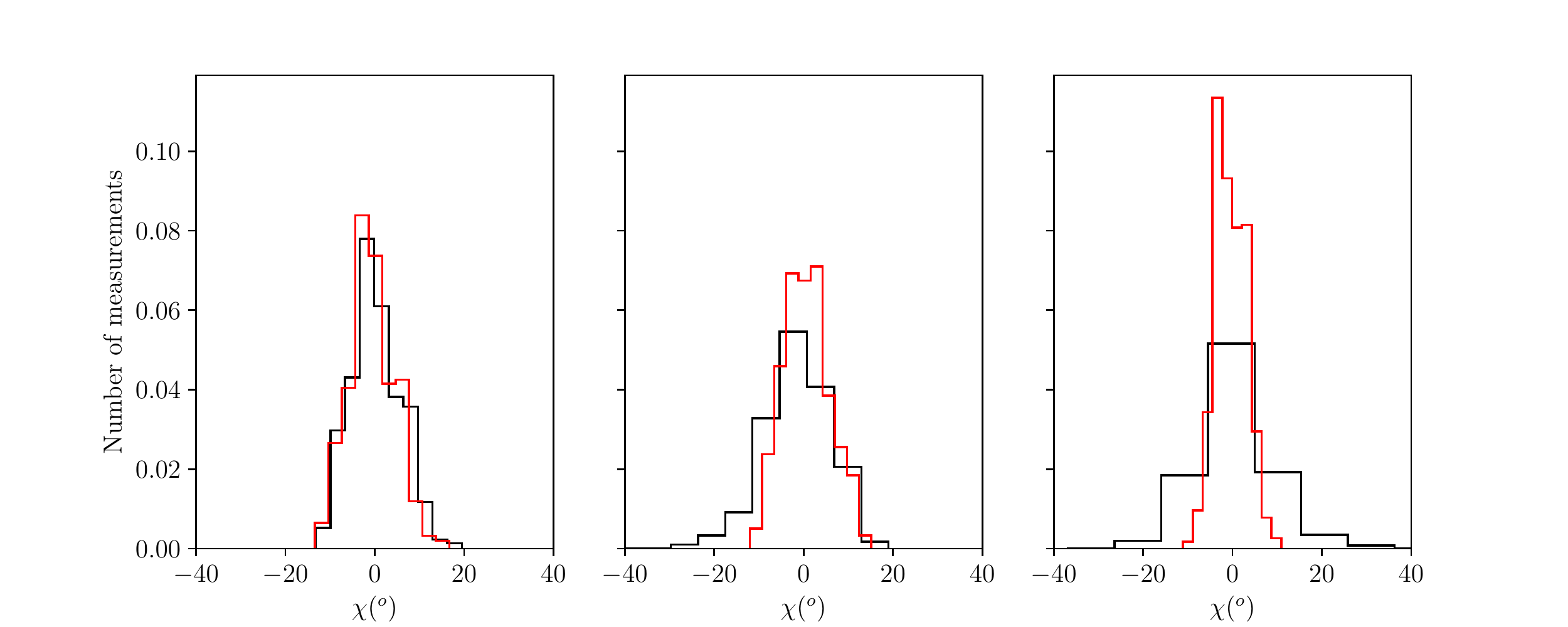}
    	\caption{Distribution of the synthetic polarization angles  for different simulation models. Black histogram corresponds to observations weighted by density, while the red one without the density weighting. \textbf{Left:} Simulation model with $M_{S}=0.7$, \textbf{Middle:} Simulation model with $M_{S}=2.0$, \textbf{Right:} Simulation model with $M_{S}=7.0$.}
    	\label{fig:pol_angle_distrib_simulations}
    \end{figure*}

\subsubsection{Projection effects and the polarization angle distribution}

Similar to velocities, fast and slow modes can also induce $\delta B/B$ variations \citep{cho_lazarian_2002}. As a result, non-Alfvénic modes will contribute to the observed signal, which will be larger than the Alfvénic. Observationaly $\delta B/B_{0}$ is computed from the spread in the polarization angle distribution, \dangles. However, \dangles\ may not trace $\delta B/B_{0}$ accurately. We present two projection effects that are related to this discrepancy. The first one has been demonstrated in previous works, while to our knowledge we are the first to demonstrate the second.

\underline{1. Degeneracy with the LOS angle}
\label{sec:degeneracy_problem}

According to equation~(\ref{eq:classical_dcf}), $B_{0}$ is inversely proportional to \dangles. This means that regions with highly disordered magnetic field, i.e. high \dangles, have a weak field strength. However, high \dangles\ can also be obtained if the magnetic field is mostly parallel to the LOS. In this case, even small perturbations will lead to POS fields that look highly disordered \citep{ostriker_2001, falceta_2008, hensley_2019}. Thus, there is a degeneracy in \dangles\ between the viewing angle of the magnetic field and the field strength \citep{ostriker_2001, falceta_2008}. \cite{houde_2004} demonstrated the appropriate geometrical modification of the DCF method in order to account for magnetic fields inclined with respect to the LOS. Such approach, however, requires knowledge of the magnetic field LOS viewing angle.

\underline{2. Non-homogeneity effect}
\label{sec:non_homogeneity_effect}

\cite{zweibel_1990} and \cite{myers_1991} proposed that the dispersion of the polarization angles, \dangles, is systematically lower due to line of sight averaging of the magnetic field directions. They argued that the polarization signal is averaged over $N$ distinct, independent regions (turbulent cells) along the LOS. Thus, \dangles\ is biased towards lower values, and the magnetic field strength is systematically overestimated. Similar conclusions on the LOS averaging were also reached by \cite{wiebe_2004}, \cite{houde_2009} and \cite{cho_2016}.

Consider, however, a Cartesian coordinate system with (discrete)  independent variables $i$, $j$, $k$. The $ij$ plane is the POS and $k$ is parallel to the LOS. The Stokes parameters are (see \cite{lee_draine_1985} and \cite{falceta_2008})
    \begin{align}
		\label{eq:stokes_parameters_I}
		& I_{ij} =  \sum_{k=1}^{L} \rho_{ijk} , \\
		& Q_{ij} = \sum_{k=1}^{L} \rho_{ijk}  \frac{  (B_{ijk}^{i})^{2} - (B_{ijk}^{j})^{2}  }{ B_{ijk}^{2} }, \\
		\label{eq:stokes_parameters_U}
		& U_{ij} = \sum_{k=1}^{L} 2\rho_{ijk}  \frac{  B_{ijk}^{i} B_{ijk}^{j}  }{ B_{ijk}^{2} },
	\end{align}
where $L$ is the LOS dimension of the cloud, $\rho_{ijk}$ the volume density of the gas, $B_{ijk}^{i}$, $B_{ijk}^{j}$ are the $i$ and $j$ component of the magnetic field respectively, and $B_{ijk}^{2}$ the square of the total field strength. The polarization angle is 
\begin{equation}
    \label{eq:pol_angle}
    \chi = 0.5 \arctan(U/Q),
\end{equation}
and the degree of polarization, 
\begin{equation}
    \label{eq:p_degree}
    p = \sqrt{Q^{2} + U^{2}}.
\end{equation}
Since the Stokes parameters are averages with density weights, density variations along the LOS may increase \dangles. We produce synthetic observations of 3D simulations in order to demonstrate this effect. Here we assume infinite resolution, which corresponds to optical polarization data. However, beam convolution should be taken into account if sub-mm data were to be simulated \citep{heitsch_2001, wiebe_2004, falceta_2008, houde_2009}.

We use the publicly available simulations from the CATS database\footnote{\url{https://www.mhdturbulence.com/}} \citep{burkhart_CATS_2020}, from which we extract the "Cho-ENO" models \citep{cho_2003_sim, burkhart_2009, portilio_208, bialy_2020}. These simulations solve the ideal MHD equations using an isothermal equation of state in a box with periodic boundary conditions and no self-gravity. Turbulence is driven sollenoidally. The simulations are scale-free in dimensionless units with the dimensionless sound speed, $\tilde{c_{s}}$, regulating the units. We convert to cgs units following \cite{hill_2008} and we adopt a sound speed $c_{s, obs}=0.91$ km/s, which is typical of $\rm{H}_{\rm{I}}$ clouds at $T=100$ K. All the models have Alfvén Mach number, defined as $M_{A}=\delta v/V_{A}$, equal to 0.7, while the sonic Mach number, $M_{s}=\delta v/\tilde{c_{s}}$, ranges from 0.7 - 7.0. We use models with $M_{A}=0.7$ only, in order to match observations, which indicate that the ISM turbulence is sub/trans-Alfvénic, see \S~\ref{sec:different_q_values} below.

We create synthetic polarization maps for every simulation model by computing the Stokes parameters and the polarization angles with equations~(\ref{eq:stokes_parameters_I})-(\ref{eq:pol_angle}). The dispersion of the polarization angles, \dangles, for each model is shown in Table~\ref{table:DCF_3D_simulation} in column 4.

In Fig.~\ref{fig:pol_angle_distrib_simulations} we show the polarization angle distributions from three different simulation setups with $M_{s}=0.7$ (left panel), $M_{s}=2.0$ (middle panel), $M_{s}=7.0$ (right panel). The black histogram corresponds to the polarization angles computed using equation~(\ref{eq:pol_angle}). The red histograms correspond to the distribution of polarization angles when we integrate by setting $\rho=1$ everywhere in the box. Histograms are normalized so that the area under each histogram integrates to one. The un-weighted distributions become narrower at larger \MS, because more independent turbulent cells are created along the LOS. On the other hand, the density-weighted distributions (black) become wider, because more significant overdensities are created due to enhanced compression at larger \MS. It appears that density fluctuations can induce extra dispersion in the observed polarization angle distribution. This result is consistent with \cite{falceta_2008}, who found that the degree of polarization decreases as \MS\ increases. Thus, in contrast to \cite{zweibel_1990} and \cite{myers_1991}, we have found that the LOS averaging of the polarization angles can induce extra dispersion and as a result the magnetic field strength is systematically underestimated. 

\begin{table*}
    \centering
    \caption{3D MHD simulation}             
    \label{table:DCF_3D_simulation}      
    \begin{tabular}{c c c c c c c c }        
    \hline\hline                 
        $M_{S}$ & $V_{A}^{true}$ & $\sigma_{v}$  & \dangles  & $V_{A}^{\rm{DCF}} (f=1)$  & 
        $ {\langle B_{t}^{2} \rangle}^{0.5} /B_{0}$  & $V_{A}^{\rm{DCF+HH09}}$ & $V_{A}^{\rm{new}}$ \\    
        \hline                        
       $0.7$ & 0.91 & 0.46 & 0.097 & 4.7  & 0.11  & 4.1 & 1.0 \\
       $1.2$ & 1.60 & 0.58 & 0.116 & 5.0  & 0.13  & 4.5 & 1.2 \\
       $2.0$ & 2.87 & 1.36 & 0.128 & 10.6  & 0.19  & 7.1 & 2.7 \\
       $4.0$ & 5.09 & 1.89 & 0.132 & 14.3  & 0.24  & 7.9 & 3.7 \\
       $7.0$ & 9.10 & 3.98 & 0.146 & 27.3 & 0.29  & 13.7 & 7.4 \\
    \hline                                   
    \end{tabular}
    \tablefoot{The following columns are measured in units of km/s: $V_{A}^{true}$, $\sigma_{v}$, $V_{A}^{\rm{DCF}}$, $V_{A}^{\rm{DCF+HH09}}$ and $V_{A}^{new}$. The quantities \dangles\ and $ \sqrt{\langle B_{t}^{2} \rangle} /B_{0}$ are measured in radians.}
\end{table*}

\subsubsection{Different $f$ values}
\label{sec:different_q_values}

\cite{chandra_fermi} assumed that \textit{turbulent motions are isotropic} and they adopted $f=1/\sqrt{3}$. If the field strength is weak, turbulent motions will drag the field lines towards random directions and turbulence will be isotropic (super-Alfvénic turbulence). 
However, there is overwhelming observational evidence that magnetic fields in the ISM have well-defined directions indicating that turbulence is sub/trans-Alfvénic, and hence turbulent properties are highly anisotropic \citep[see for example,][]{montgomery_1981, shebalin_1983, higdon_1984, sridhar_1994, goldreich_1995, goldreich_1997}.  \cite{heyer_2008} using CO data, found that velocity structures in Taurus are highly anisotropic. In the same region, \cite{goldsmith_2008} reported the existence of highly-anisotropic density structures, which are aligned parallel to the mean field, known as striations. Striations have also been observed in the Polaris Flare \citep{panopoulou_2015} and Musca \citep{cox_2016, tritsis_2018} and they are formed due to magnetosonic waves \citep{tritsis_2016} in sub-Alfvénic turbulence \citep{beattie_2020_striations}. More evidence for ordered magnetic fields in molecular clouds can be found in \cite{franco_2010}, \cite{franco_2015}, \cite{pillai_2015}, \cite{hoq_2017}, and \cite{tang_2019}. \cite{stephens_2011} explored the magnetic field properties of 52 star forming regions in our Galaxy and concluded that more than $80\%$ of their targets exhibit ordered magnetic fields. The diffuse atomic clouds in our Galaxy are preferentially aligned with the magnetic field \citep{clark_2014} implying the importance of the magnetic field in their formation. \cite{planck_collaboration_2016} studied a larger sample of molecular clouds in the Goult Belt and concluded that density structures align parallel or perpendicular to the local mean field direction. This is also consistent with sub/trans-Alfvfénic turbulence, e.g. \cite{soler_2013}. In addition, \cite{mouschovias_2006}, using Zeeman data, concluded that turbulence in molecular clouds is slightly sub-Alfvénic as well. All these lines of evidence indicate that ISM turbulence is sub/trans-Alfvénic, and hence anisotropic. 

Other $f$ values were proposed when the DCF method was applied in numerical simulations. \cite{ostriker_2001} performed MHD numerical simulations of giant molecular clouds. They produced synthetic observations and suggested that the DCF equation with $f=0.5$ produces accurate measurements for their sub-Alfvénic model. \cite{padoan_2001} simulated protostellar cores and tested the DCF equation in three different cores in a super-Alfvénic MHD turbulent box. They varied the position of the observer with respect to the magnetic field direction and found that on average $f=0.4$. However, note that their values range from $0.29$ up to $0.74$ (see their Table 1). \cite{heitsch_2001} performed 3D MHD simulations of molecular clouds. They found that $f$ lies in the interval $0.33 - 0.5$ (see their Figure~6) in their three models with strong magnetic fields (sub-Alfvénic turbulence). 

In these works $f$ was found to vary significantly, but a value of $f=0.5$ is widely used, e.g. \citet{pattle_2019}. However, no physical connection of $f$ with the turbulent properties of the medium or with specific LOS averaging effects has been demonstrated. Thus, it remains unclear, which value of $f$ is most appropriate for any given real physical cloud. The number of turbulent eddies along the line of sight may be a relevant metric of $f$ and can be estimated with the HH09 method or following the approach of \cite{cho_2016} and \cite{yoon_cho_2019}.

\subsection{Testing classical DCF with 3D simulations}

We test the DCF method in the 3D numerical simulation we used in \S~\ref{sec:non_homogeneity_effect}. We create synthetic spectroscopic data following \cite{miville_2003}, who assumed optically thin emission and included no chemistry. We set the LOS parallel to the $z$-axis. We compute the PPV cube, $I_{v}(x, y, v)$, along the LOS using the following equation,
\begin{equation}
    \label{eq:velocity_cubes}
    I_{v}(x, y, v) = \sum_{z} \frac{\rho(x, y, z) \delta z}{\sqrt{2\pi} \sigma(x, y, z)} {\rm exp} \left [ -\frac{ \left ( v_{los}(x, y, z) - v \right )^{2} } {2\sigma(x, y, z)^{2}} \right ],
\end{equation}
where $v_{los}(x, y, z)$ is the LOS velocity component, $v$ is the central velocity of each velocity channel and $\delta z =1$ pixel. The velocity spread is
\begin{equation}
    \sigma(x, y, z) = \left [ \left (\frac{\partial v_{los}(x, y, z)}{\partial z} \delta z \right )^{2} + \frac{k_{B}T}{m} \right ]^{1/2}.
\end{equation}

   \begin{figure}
   \centering
   \includegraphics[width=\hsize]{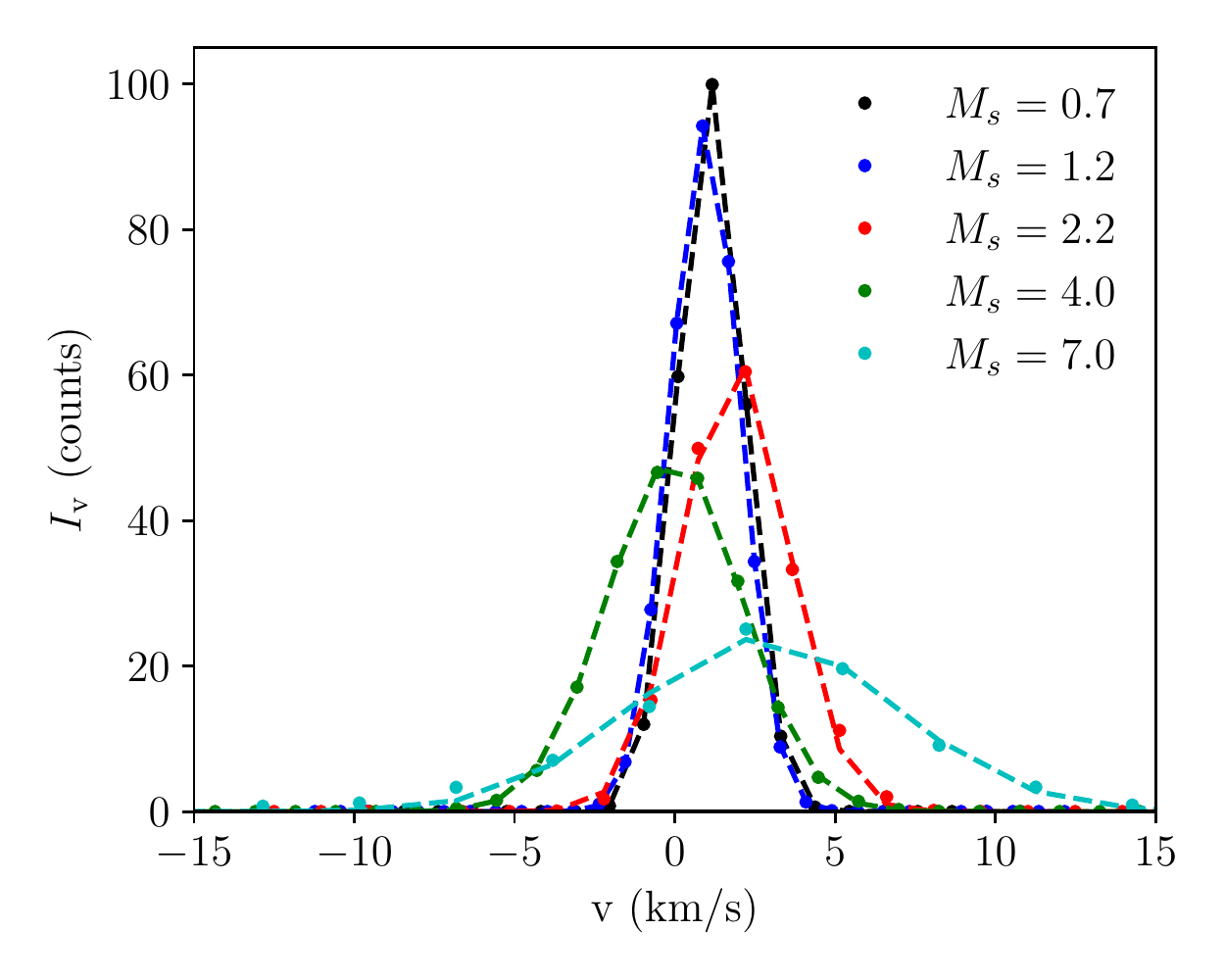}
      \caption{Synthetic emission line profiles for different simulation models. All the models have Alfvén Mach number $M_{A}=0.7$. The sonic Mach number, \MS, for each simulation model is shown in the legend. The dashed lines correspond to the fitted Gaussian profiles.}
    \label{fig:spectral_lines_simulation}
   \end{figure}

In Fig.~\ref{fig:spectral_lines_simulation} we show the mean emission profile of each simulation setup derived using equation~(\ref{eq:velocity_cubes}). The dots represent the intensity, $I_{v}$, versus the gas velocity and the dashed lines the fitted Gaussians. From the fitted Gaussians we derive the standard deviation of each profile, shown in the $\sigma_{v}$ column of Table~\ref{table:DCF_3D_simulation}. In the same Table we show the standard deviation of the polarization angle distribution, \dangles, for each setup, and the estimated Alfvén speed, using the DCF relation (equation~\ref{eq:classical_dcf} with $f=1$) in the $V_{A}^{\rm{DCF}}$ column. The true Alfvén speed for each setup is shown in the column $V_{A}^{true}$. 

The DCF method systematically overestimates the magnetic field strength. For models with $M_{s}=0.7$ DCF without an $f$ factor ($f=1$) deviates from the true value by a factor of $\sim 5$. This implies that $f=1/5$ in order for the method to produce accurate estimates. On the other hand, for models with $M_{s}=1.2, 4.0$  and $7.0$ the correction factor has to be $f \sim 1/3$. It is evident that the generally adopted $f=0.5$ (\S~\ref{sec:different_q_values}) value does not apply for these models. Our results indicate that $f$ decreases with  \MS, but it remains to be shown whether and why this can be considered a general property.

\section{HH09 method }
\label{sec:hdcf_method}

\subsection{Foundations of the method}

An alternative way to estimate the $\delta B /B_{0}$ ratio has been presented by \cite{hildebrand_2009} and \cite{houde_2009}, here HH09. The method was developed in order to avoid inaccurate estimates of the magnetic field strength induced by sources other than MHD waves, e.g. large-scale bending of the magnetic field due to gravity and differential rotation. HH09 computed the \textit{isotropic dispersion function} of the polarization map as
\begin{equation}
    \label{eq:sf_observational}
    \left\langle \rm{cos} \left [ \Delta \phi(l) \right ]   \right\rangle =  \langle \rm{cos} \left [ \Phi( \vec{x}) - \Phi(\vec{x}+\vec{l)} \right ] \rangle .
\end{equation}
The quantity $\Phi(x)$ denotes the polarization angle measured in radians, $\vec{x}$ denotes the 2D coordinates in the POS , $\vec{l}$ the spatial separation of two polarization measurements in the POS and brackets averaging over the entire polarization map. The polarization angle differences are constrained in the interval $[0\degr, 90\degr]$. 

HH09 defined the total magnetic field as $\vec{B}_{tot}=\vec{B_{0}}+\vec{B_{t}}$, where \Bordered\ is the mean magnetic field component and \Bturb\ the turbulent (or random) component. HH09 assumed that the strength of \Bordered\ is uniform and \Bturb\ is induced by gas turbulent motions. They derived the following analytical relation for equation~(\ref{eq:sf_observational}),
\begin{equation}
\label{eq:hildebrand_houde_2009}
1 - \langle \rm{cos} \left [ \Delta \phi (l) \right ] \rangle  \simeq  \sqrt{2\pi} \frac{\langle B_{t}^{2} \rangle}{B_{0}^{2}} \frac{\delta^{3}}{(\delta^{2}+2W^{2})\Delta'} \times (1 - e^{l^{2}/2(\delta^{2} + 2W^{2})}) + ml^{2},
\end{equation}
where $m$ is a constant, $\Delta'$ is the effective cloud depth and $W$ the beam size. The effective cloud depth is always smaller than the size of the cloud ($L$), $\Delta ' \leq L$, and is defined as the FWHM of the auto-correlation function of the polarized intensity \citep{houde_2009}. The validity of this relation is limited to spatial scales $ \delta \leq  l \leq d$, where $\delta$ is the correlation length of \Bturb\ and $d$ is the upper limit below which \Bordered\ remains uniform. This equation is used to estimate the $\langle B_{t}^{2} \rangle^{1/2}/B_{0}$ term, which is then inserted in the DCF formula as,
\begin{equation}
    \label{eq:Va_hh09}
    V^{\rm{DCF+HH09}}_{A} \simeq \sigma_{v} \Bigg [\frac{\langle B_{t}^{2} \rangle^{1/2}}{B_{0}} \Bigg]^{-1}.
\end{equation}
The only difference to the classical DCF is that the $\delta B/B_{0}$  term is obtained from the fit of equation~(\ref{eq:hildebrand_houde_2009}) instead of the dispersion of the polarization angle distribution.

In order to use the method, one has to compute the dispersion function using equation~(\ref{eq:sf_observational}), and then fit the model in the right hand side of equation~(\ref{eq:hildebrand_houde_2009}). The fit has the following three free parameters: $\langle B_{t}^{2} \rangle / B_{0}^{2}$, $\delta$ and $m$.

\subsection{Caveats of the method}
\label{sec:hh09_critisicm}

\subsubsection{Omission of the \texorpdfstring{$\vec{\delta B \cdot \vec{B_{0}}}$}{TEXT} term}
\label{sec:omission_BtB0}

\cite{hildebrand_2009} assumed that \textit{the correlation of $B_{t}$ and $B_{0}$ is zero, i.e. $\langle B_{0} \cdot B_{t} \rangle = 0$}, where the averaging is over the full map (see their equation A2). We identify three different regimes, in which $\langle \vec{\delta B} \cdot \vec{B_{0}} \rangle = 0$ \footnote{Note that we use the more general notation $\vec{\delta B}$ instead of $\vec{B_{t}}$ since it refers to the fluctuating component of the magnetic field not necessarily in turbulent conditions.}:
\begin{enumerate}
    \item \underline{Super-Alfvénic turbulence}\\
    In the highly super-Alfvénic regime, MHD turbulence behaves like hydro turbulence and $\vec{\delta B}$ is random, i.e. $\langle \vec{\delta B} \rangle = 0$. In this case \Bordered\ is much weaker than $\vec{\delta B}$ and the two quantities are statistically independent. Thus, 
    \begin{equation}
        \langle \vec{\delta B} \cdot \vec{B_{0}} \rangle = \langle \vec{\delta B} \rangle \cdot \langle \vec{B_{0}} \rangle = 0.
    \end{equation}
    \item \underline{Purely Alfénic (or incompressible) turbulence}\\
    For Alfvénic turbulence $\vec{\delta B} \cdot \vec{B_{0}}=0$, because Alfvén waves are transverse and their field fluctuations, $\vec{\delta B}$, are always perpendicular to the mean field $\vec{B_{0}}$ \citep{goldreich_1995}. 
    \item \underline{Force-free field}\\
    If we use the linearized ($|\vec{\delta B}| \ll |\vec{B_{0}}|$) induction equation,
    \begin{equation}
        \label{eq:induction_equation}
        \vec{\delta B} = \vec{\nabla} \times (\vec{\xi} \times  \vec{B_{0}}),
    \end{equation}
    where $\vec{\xi}$ denotes the gas displacements vector, it can be shown that \citep{spruit_2013}
    \begin{equation}
        \label{eq:DeltaEpsilon}
        \langle \vec{\delta B} \cdot \vec{B_{0}} \rangle = - \frac{1}{4\pi} \int \vec{\xi} \cdot [(\vec{\nabla} \times \vec{B_{0}}) \times \vec{B_{0}} ]  dV .
    \end{equation}
    If the field is force-free, i.e. $(\vec{\nabla} \times \vec{B_{0}}) \times \vec{B_{0}}=0$, the above equation implies that $\langle \vec{\delta B} \cdot \vec{B_{0}} \rangle=0$. Although it is occasionally used as an approximation a force-free field naturally decays without causing fluid motions \citep{chandra_kendall}. 
\end{enumerate}

The cross term, $\langle \vec{\delta B} \cdot \vec{B_{0}} \rangle$, is connected with the compressible modes \citep{montgomery_1987}, for which $ \vec{\delta B} \cdot \vec{B_{0}} \neq 0$ \citep{goldreich_1995}. \cite{Bhattacharjee_1988} and \cite{Bhattacharjee_1998} using standard perturbation theory derived the first order solution of the equation of motion
\begin{equation}
    \label{eq:magnetostatic}
    \delta P + \vec{\delta B} \cdot \vec{B_{0}} = 0,
\end{equation}
where $\delta P$\footnote{Note that \cite{Bhattacharjee_1988} and \cite{Bhattacharjee_1998} use the subscript $1$ instead of $\delta$ for the perturbed quantities.} is the gas-pressure perturbations. The equation of state connects gas density and pressure, and hence
\begin{equation}
    \label{eq:coupling}
    \delta \rho \propto \vec{\delta B} \cdot \vec{B_{0}}
\end{equation}
where $\delta \rho$ are the gas density perturbations. Equation~(\ref{eq:coupling}) explicitly shows the coupling between $\delta \rho$ and $\vec{\delta B} \cdot \vec{B_{0}}$. Hence, $\langle \vec{\delta B} \cdot \vec{B_{0}} \rangle \propto \langle \delta \rho \rangle$, which implies that $\langle \vec{\delta B} \cdot \vec{B_{0}} \rangle = 0$ only if $\langle \delta \rho \rangle =0$, which is true for purely incompressible turbulence, generally not realizable in the ISM. The omission of the $\vec{\delta B} \cdot B_{0}$ term may lead to significant inaccuracies of the estimate of $\langle B_{t}^{2} \rangle^{1/2}/B_{0}$ when the HH09 method is applied in sub/trans-Alfvénic and compressible turbulence.

\subsubsection{Isotropic turbulence}
HH09 assumed that \textit{turbulence is isotropic} and they used a global correlation length, $\delta$. As we argued in \S~\ref{sec:different_q_values}, this is against observational evidence, which shows highly anisotropic structures and properties. Anisotropic media exhibit different correlation lengths perpendicular and parallel to the mean field. This has been shown in the observations of \cite{higdon_1984} and \cite{heyer_2008} and by numerous theoretical works, e.g. \cite{shebalin_1983}, \cite{goldreich_1995}, \cite{sridhar_1994}, \cite{cho_vishniac_2000}, and \cite{maron_2001} \citep[see][for a recent review]{oughton_2020}. This indicates that anisotropic structure functions, similar to \cite{cho_vishniac_2000} and \cite{maron_2001}, should be used instead. \cite{Chitsazzadeh_2012} and \cite{houde_2013} refined the HH09 method and considered the anisotropic turbulent properties of sub/trans-Alfvénic turbulence. Our criticism, however, focuses on the original isotropic version of the method, which is still often applied, e.g. \cite{chuss_2019}.

\subsubsection{Sparsity of data and the HH09 method}

The computation of the dispersion function (equation~\ref{eq:sf_observational}) for sparsely sampled data introduces a bias. \cite{soler_2016} showed that the dispersion function computed with optical polarization is slightly different than the one computed with sub-mm data, see their Fig.~6. They commented that the sparse sampling of the optical polarization measurements induces a "jittering". Their figures indicate that the structure function with the optical data systematically overestimates the intercept of the function. As a result, for the case of the sparse sampling, the $\langle B_{t}^{2} \rangle^{1/2}/B_{0}$ parameter would be biased towards higher values. No solution to this problem has been suggested yet.

\subsection{Testing HH09 with 3D simulations}

\begin{figure}
	\includegraphics[width=\hsize]{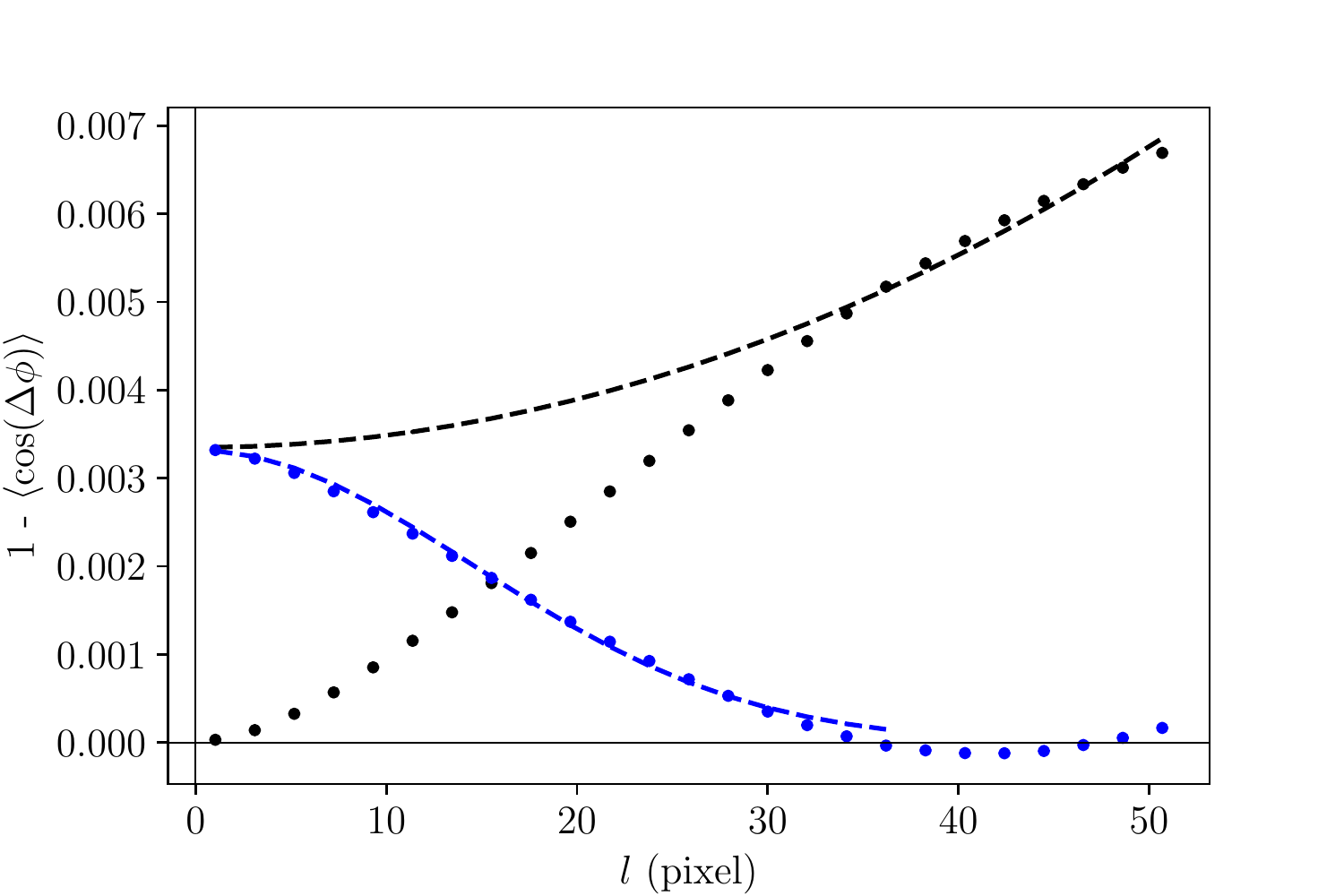}
	\caption{Black dots correspond to the dispersion function computed using equation~(\ref{eq:sf_observational}) for the $M_{s}=0.7$ model. The model fit at large scales is shown with the black broken curve. Blue points mark the dispersion function subtracted with the black broken line and the blue line shows the fit of equation~(\ref{eq:autocorrelation_function}).}
	\label{fig:structure_function_ms_07}
\end{figure}

We apply the HH09 method to the polarization angle maps, $\chi(x, y)$, we created in \S~\ref{sec:non_homogeneity_effect} as suggested by \cite{houde_2009}. We compute the dispersion function (equation~\ref{eq:sf_observational}) from our synthetic data, black dots in Fig.~\ref{fig:structure_function_ms_07}. We then fit to the black dots the right hand side of equation~(\ref{eq:hildebrand_houde_2009}), without including the exponential term. This fit, shown with the black broken curve, is performed at larger scales, i.e $l > 20$ pixels. The black broken curve is then subtracted from the black points and the result, shown with the blue dots, corresponds to the turbulent auto-correlation function, i.e. to the second term of equation~(\ref{eq:hildebrand_houde_2009})
\begin{equation}
    \label{eq:autocorrelation_function}
    b^{2}(l) = \sqrt{2\pi} \frac{\langle B_{t}^{2} \rangle}{B_{0}^{2}} \frac{\delta^{3}}{(\delta^{2}+2W^{2})\Delta'}  e^{l^{2}/2(\delta^{2} + 2W^{2})}.
\end{equation}
We fit this term to the blue points and we derive the turbulent correlation length (blue broken curve in Fig.~\ref{fig:structure_function_ms_07}), $\delta$. The turbulent-to-ordered magnetic field ratio is 
\begin{equation}
    \label{eq:btbo}
    \frac{\langle B_{t}^{2} \rangle^{1/2}}{B_{0}} = b(0) \sqrt{N},
\end{equation}
where $N$ is the number of turbulent cells along the line of sight and is defined as,
\begin{equation}
    N = \frac{\delta^{2} + 2W^{2} }{\sqrt{2 \pi} \delta ^{3}} \Delta',
\end{equation}
where $\Delta '$ is computed as in \cite{houde_2009}. Like in optical polarization data beam resolution is infinite in our synthetic observations, and hence $W=0$ \citep{franco_2010, panopoulou_2016}. In Appendix~\ref{sec:sf_fitting} we show the dispersion function plots for the rest of the models and in Table~\ref{table:hh09_fitting_parameters} the best fit parameters are given.

The turbulent-to-ordered ratio obtained for the different simulation models is shown in column ${\langle B_{t}^{2} \rangle}^{1/2}/B_{0}$, in Table~\ref{table:DCF_3D_simulation}. The estimated Alfvén speed when the HH09 method is combined with DCF is shown in the same Table in the $V_{A}^{\rm{DCF+HH09}}$ column. The DCF+HH09 estimates are significantly improved compared to the classical DCF values although the overestimation from the true Alfvén speed ($V_{A}^{true}$) is still prominent. We note, however, that if we consider that the effective cloud depth is equal to the cloud size, i.e. $\Delta '=256$ pixels, the HH09 method produces more accurate estimates of the field strength for the models with $M_{s}= 4.0$ and $7.0$.

\section{Proposed method}
\label{sec:proposed_method}

Motivated by the existence of compressible modes and the contamination they induce in the aforementioned methods, we propose a generalized method, which takes into account these modes. We start with equation~(\ref{eq:magnetic_energy}). Similar to DCF, we assume that gas is perfectly attached to the magnetic field and that turbulent motions are completely transferred to magnetic fluctuations. Unlike  DCF method, we assume \textit{that all MHD modes are excited, including fast and slow modes}. Since we include the compressible modes, we do not omit the cross-term in the magnetic energy, which for the general case is $\vec{\delta B} \cdot \vec{B_{0}} \neq 0$ (see \S~\ref{sec:omission_BtB0}). To leading order, the fluctuating part in the energy equation, when $|\vec{\delta B}| \ll |\vec{B_{0}}|$, will be \citep{federrath2016}
\begin{equation}
    \label{eq:fluct_energy_eq}
    \delta \epsilon_{m} \simeq  \frac{\delta B B_{0}}{4 \pi}.
\end{equation}
The above equation implies that $\vec{\delta B}$ is aligned to $\vec{B_{0}}$. This can be understood due to the local alignment between $\vec{\delta B}$ and $\vec{\delta v}$, which appears in sub-Alfvénic turbulence \citep{boldyrev_2005, boldyrev_2006}. This is known as \textit{dynamic alignment} and has been observed in numerical simulations of both incompressible, e.g. \cite{mason_2006}, and compressible, e.g. \cite{Kritsuk_2017}, turbulence. Velocity gradients parallel to $\vec{B_{0}}$, which only appear in compressible turbulence, drag $\vec{\delta B}$ to align with $\vec{B_{0}}$ \citep{beattie_2020} due to the dynamic alignment. Despite its simplicity, the accuracy of equation~(\ref{eq:fluct_energy_eq}) was proven to be remarkable when tested in numerical simulations of sub-Alfvénic and compressible turbulence \citep{federrath2016, beattie_2020}. We assume that turbulent kinetic energy is equal to magnetic energy fluctuations, hence
\begin{equation}
    \frac{1}{2} \rho \delta v^{2} = \frac{\delta B B_{0}}{4 \pi} 
\end{equation}
As in DCF, we use the relation $\delta \theta = \delta B/B_{0}$, since the turbulent-to-ordered ratio is measured from the polarization angle distribution. We obtain
\begin{equation}
    \label{eq:new_method_relation}
    B_{0} = \sqrt{2 \pi \rho}\frac{\delta v}{\sqrt{\delta \theta}} , 
\end{equation}
which can be solved for the Alfvén speed, $V_{A}=B_{0}/\sqrt{4\pi \rho}$, as
\begin{equation}
    \label{eq:new_method_speed}
    V_{A} = \frac{1}{\sqrt{2}}\frac{\delta v}{\sqrt{\delta \theta}}.
\end{equation}

As in DCF, we assume that there is a guiding field about which MHD waves propagate. We also assume that turbulent kinetic energy is equal to the fluctuating magnetic energy, \textit{which is given by $\delta B B_{0}$, since it is the dominant (first order) term in the magnetic energy.} This term is neglected in previous methods. The major difference between our equation and the DCF is that the $(\delta B/B_{0})^{n}$ term in our equation appears with $n=-1/2$, while in DCF it is $n=-1$. 

\subsection{Testing the proposed method with 3D simulations}

We apply the proposed method to the 3D compressible MHD simulations, see \S~\ref{sec:non_homogeneity_effect}, in order to test its validity. We use the $\delta \theta$ and $\sigma_{v}$ from Table~\ref{table:DCF_3D_simulation}. In the same Table we show in the $V_{A}^{new}$ column the value computed using equation (\ref{eq:new_method_speed}). It is evident that our model produces acceptably accurate estimates of the true \VA\ in all models. 

\begin{figure}
  \centering
   \includegraphics[width=\hsize]{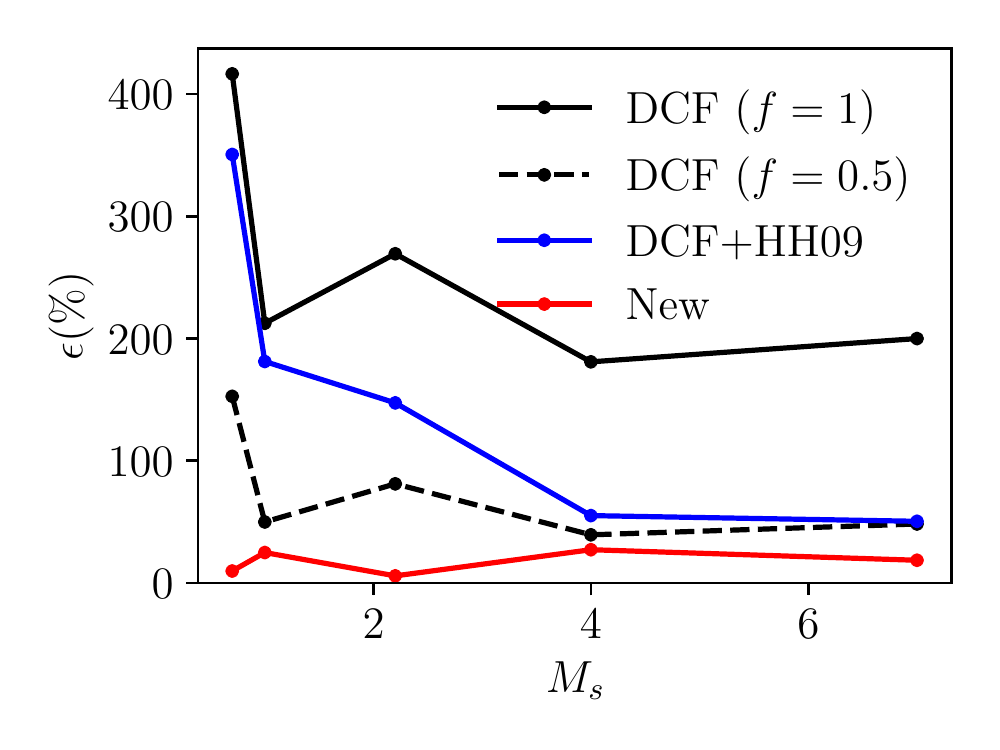}
      \caption{Relative deviation of each method estimate for the different simulation models. All the models have $M_{A}=0.7$.}
    \label{fig:summary}
\end{figure}

The relative deviation of the estimated \VA\ from the true one is 
\begin{equation}
    \epsilon = \frac{|V_{A}^{true} - V_{A}^{est}|}{V_{A}^{true}},
\end{equation}
where $V_{A}^{true}$ is the true \VA\ and $V_{A}^{est}$ is the estimated value from the various methods. In Fig.~\ref{fig:summary} we show $\epsilon (\%)$ for all models. The black points correspond to DCF with $f=1$ (solid line) and $f=0.5$ (dotted line), blue to HH09, and red to the new proposed method (equation~\ref{eq:new_method_speed}). DCF systematically overestimates the true value and even with the previously reported two-fold reduction, i.e. $f=0.5$, the deviations are still significant. DCF+ HH09 produces very large estimates in the models with low \MS, while at higher \MS\ the method is more accurate. Our proposed method produces quite accurate values for the field strength independently of \MS. The mean deviation, $\epsilon$, of our method is $17 \%$. The largest deviation is seen in the model with $M_{s}=4.0$, where $\epsilon = 27\% $.

The  accuracy of our method does not depend on \MS. However our method systematically underestimates the true field strength. This can be explained by the non-homogeneity effect that we discussed in \S~\ref{sec:non_homogeneity_effect}. The polarization map includes dispersion due to density enhancements, hence \dangles\ overestimates the $\delta B/B_{0}$ ratio. Only in the model with $M_{s}=0.7$, where density inhomogeneities are minimal, the method slightly overestimates the field strength. 

\subsection{Limitations of our method}
There are limitations of our method related to the observables and the physics behind the proposed equation.

\begin{enumerate}
    \item Polarization angle maps are affected by density inhomogeneities, hence they do not probe the $\delta B/B_{0}$ term accurately. The observed spread is higher than the true ratio, hence the field strength is underestimated. This non-homogeneity effect is the dominant uncertainty in our method.
    \item As in DCF, we neglect the contribution of self-gravity. In the presence of self-gravity the field lines are bent and the observed \dangles\ and $\delta v$ are higher. Other methods have to be employed prior to ours, e.g. \cite{girart_2006} and \cite{pattle_2017}, in order to remove the self-gravity effects from both polarization and spectroscopic maps.
    \item Not all the MHD modes induce $\delta B/B_{0}$ variations. For example, for very strong fields, fast modes can propagate perpendicularly to the mean field and oscillate like an harmonica, without disturbing the field lines. These modes will contribute in velocity, but not in the polarization angle, biasing the magnetic field estimate to higher values. The tests we performed though show that this is not a very important effect.
\end{enumerate}

\section{Discussion \& Conclusions}
\label{sec:conclusions}

Dust polarization traces the magnetic field orientation in the POS, but not its strength. DCF and HH09 are the most widely applied methods that employ polarization data in order to estimate the strength of the field. They rely on the assumption that isotropic gas turbulent motions induce the propagation of small amplitude Alfvén waves, $|\vec{\delta B}| \ll |\vec{B_{0}}|$. Observations indicate that turbulence in the ISM is highly anisotropic, e.g. \cite{higdon_1984} \cite{heyer_2008} and \cite{planck_collaboration_2016}, see also \S~\ref{sec:different_q_values}. The sufficiently high \MS\ in the ISM, e.g. \citep{heiles_2003}, implies that compressible modes are important. Both DCF and HH09 neglect the compressible modes. As a result, the estimates of the magnetic field they provide may deviate significantly from the true value. 

We tested the DCF method in synthetic observations from publicly available 3D MHD simulations. We found that the DCF method systematically overestimates the true strength value. The previously reported reduction factor $f=0.5$ is not supported by our analysis. We found similar results when we tested the HH09 method. However, the accuracy of the latter improves significantly for models with high sound Mach numbers. 

We proposed a new method to estimate the magnetic field strength in the ISM. We accounted for all MHD modes, both Alfvén and compressible. We assumed that: wave fluctuations are sufficiently small compared to the mean field strength, $|\vec{\delta B}| \ll |\vec{B_{0}}|$;  turbulent kinetic energy is equal to the fluctuating part of the magnetic energy density; the fluctuating magnetic energy is dominated by the $\delta B B_{0}$ cross-term, which represents the compressible modes. Instead of 
\begin{equation}
    B_{0} = f \sqrt{4\pi \rho} \frac{\delta v}{\delta \theta},
\end{equation}
we proposed the following equation:
\begin{equation}
    B_{0} = \sqrt{2 \pi \rho} \frac{\delta v}{\sqrt{\delta \theta}},
\end{equation}
where $\rho$ is the gas density, $\delta v$ is the spread in spectroscopic data emission lines and \dangles\ is the dispersion of polarization angles.

We tested the validity of our equation in synthetic observations and we compared its accuracy to that of classical DCF and DCF combined with HH09. Our method outperforms these methods and it guarantees a uniformly low error independent of Mach number and \textit{without the need for a correction factor.}

\begin{acknowledgements}
This work is dedicated to the memory of Roger Hildebrand, a pioneer in this field and an inspiring mentor. We are grateful to the referee M. Houde for his careful and constructive reviewing. RS would like to thank J.Beattie and A. Tsouros for productive discussions and Dr. B. Burkhart for her help with their numerical simulations. We thank A. Tritsis for his invaluable comments. We also thank N. D. Kylafis for his careful reading of the manuscript, V. Pavlidou and G. V. Panopoulou for fruitful discussions. RS would like to thank Dr. K. Christidis for his constant support during this project. This project has received funding from the European Research Council (ERC) under the European Unions Horizon 2020 research and innovation programme under grant agreement No. 771282.
\end{acknowledgements}

%
%

\bibliographystyle{aa}
\bibliography{bibliography}
\label{sec:structure_function_fitting}

\begin{appendix}


\begin{figure}
  \centering
   \includegraphics[width=\textwidth]{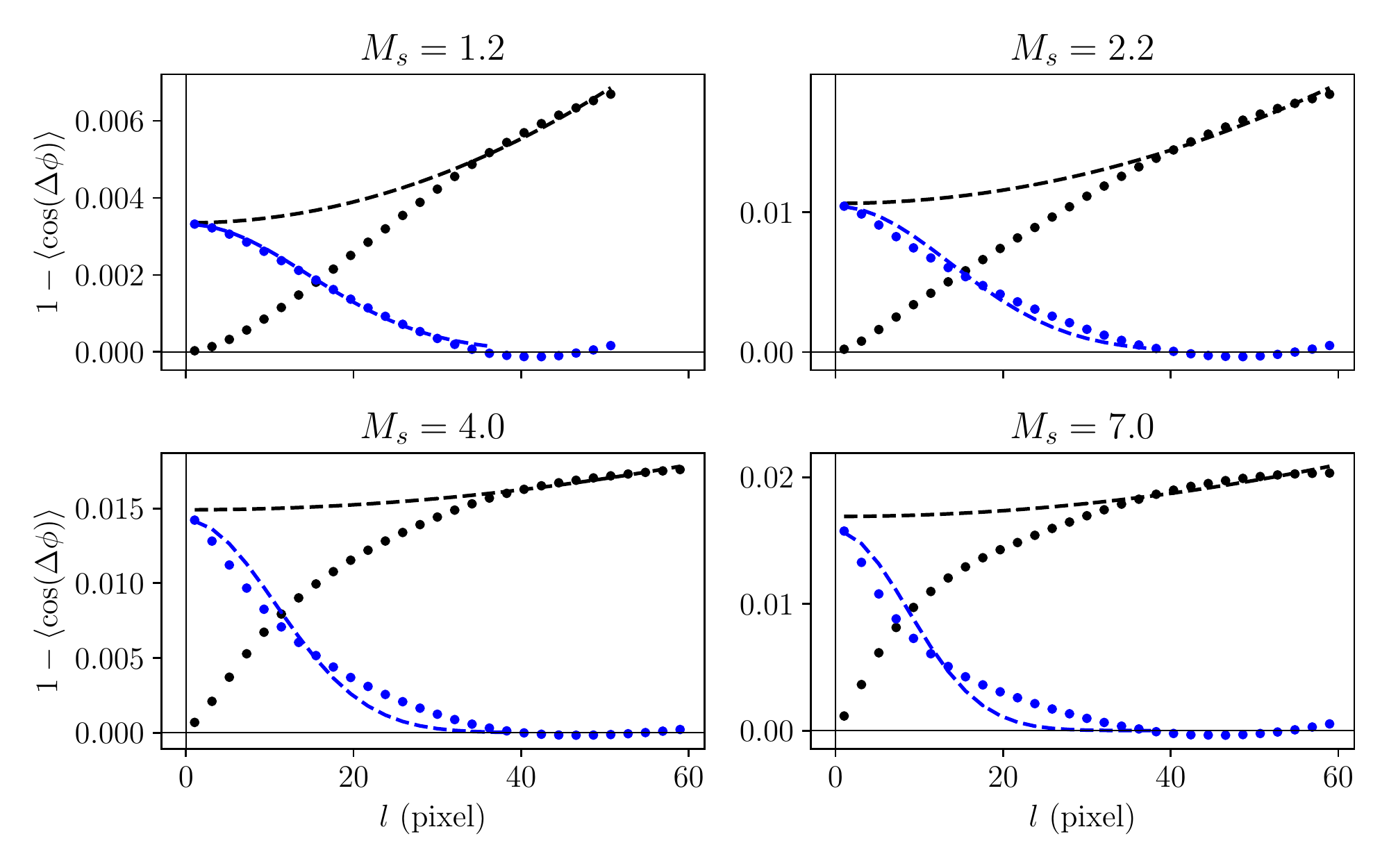}
      \caption{As in Fig.~\ref{fig:structure_function_ms_07} for the simulation models with $M_{s}=1.2-7.0$.}
    \label{fig:structure_functions}
\end{figure}

\begin{table}
    \centering
    \caption{Fitting parameters of the HH09 method.}             
    \label{table:hh09_fitting_parameters}      
    \begin{tabular}{c c c c c c}        
    \hline\hline                 
    \vspace{0.1cm}
    $M_{s}$ & $\delta$ (pixel) & $m$ & $b(0)$ & $\Delta'$ (pixels) & $N$\\    
    \hline                        
       0.7 & 15.5 & 0.0012 & 0.06 & 117 & 3.5 \\      
       1.2 & 13.7 & 0.0016 & 0.07 & 117 & 3.4  \\
       2.2 & 13.1 & 0.0015 & 0.10 & 117 & 3.6  \\
       4.0 & 10.3 & 0.0009 & 0.12 & 108 & 4.2 \\
       7.0 & 8.6  & 0.0011 & 0.13 & 112 & 5.2 \\ 
    \hline                                   
    \end{tabular}
    \tablefoot{b(0) is computed from equation~(\ref{eq:hildebrand_houde_2009}).}
\end{table}

\section{Dispersion function fitting}
\label{sec:sf_fitting}

The dispersion function fits for the models with $M_{s}=1.2 - 7.0$ are shown in Fig.~\ref{fig:structure_functions}. The best fit parameters for every model is given in Table~\ref{table:hh09_fitting_parameters}.

\end{appendix}
\end{document}